\documentclass[preprint,showpacs,preprintnumbers,amsmath,amssymb,11pt]{revtex4}
\usepackage{graphicx}
\usepackage{dcolumn}
\usepackage{bm}
\begin{document}
\title{Constraints from CMB in the Intermediate Brans-Dicke Inflation}
\author{Antonella Cid}
\email{acidm@ubiobio.cl}
\affiliation{Departamento de F\'isica, Facultad de Ciencias, Universidad del B\'io-B\'io,
Avenida Collao 1202, Casilla 5-C, Concepci\'on, Chile}
\author{Sergio del Campo}
\email{sdelcamp@ucv.cl}
\affiliation{ Instituto de F\'{\i}sica, Pontificia Universidad Cat\'{o}lica de Valpara\'{\i}so, Casilla 4059, Valpara\'{\i}so, Chile.}
\date{\today}
\begin{abstract}
We study an intermediate inflationary stage in a Jordan-Brans-Dicke
theory. In this scenario we analyze the quantum fluctuations
corresponding to adiabatic and isocurvature modes. Our model is
compared to that described by using the intermediate model in
Einstein general relativity theory. We assess the status of this
model in light of the seven-year WMAP data.
\end{abstract}
\maketitle
\section{Introduction}

The inflationary paradigm \cite{inflation,Linde} has been confirmed
as the most successful candidate for explaining the physics of the
very early universe \cite{KKMR}. This sort of scenarios has been
successful in solving some of the puzzles of the standard
cosmological model, such as the horizon, flatness and entropy
problems, as well as providing for a mechanism to seed structure
in the universe.

The source of inflation is a scalar field (the inflaton field),
which plays an important role in providing a causal interpretation
of the origin of the observed anisotropy of the cosmic microwave
background radiation, and also distribution of large scale
structure \cite{astro,astro2}. The nature of this scalar field
may be found by considering one of the extensions of the standard
model of particle physics based on grand unified theories,
supergravity, or some effective theory at low dimension of a more
fundamental string theory.

The idea that inflation comes from an effective theory is in
itself very appealing. The main motivation to study this sort of
model comes from string/M-theory. This theory suggests that in
order to have a ghost-free action high order curvature invariant
corrections to the Einstein-Hilbert action must be proportional to
the Gauss-Bonnet (GB) term \cite{BD}. GB terms arise naturally as
the leading order of the $\alpha$ expansion to the low-energy
string effective action, where $\alpha$ is the inverse string
tension \cite{KM}. This kind of theory has been applied to the
possible resolution of the initial singularity problem
\cite{ART}, also to the study of black-hole solutions \cite{
Varios1}, and accelerated cosmological solutions \cite{
Varios2}. In particular, it has been found {\cite{Sanyal}} that
for a dark energy model the GB interaction in four dimensions with
a dynamical dilatonic scalar field coupling leads to a solution of
the form $a(t) = a_0 \exp({A t^{f}})$,  where the universe starts
evolving with a decelerated exponential expansion. Here, the
constant $A$ is given by $A= \frac{2}{\kappa n}$ and
$f=\frac{1}{2}$, with $\kappa^2 = 8 \pi G$, where $G$ is the newtonian gravitational constant, and $n$ an arbitrary constant.

We have the particular scenario called intermediate
inflation \cite{Barrow1}, characterized for the scale factor
evolving as $a(t)=a_0\exp(A t^f)$. In this model, the expansion of the
universe is slower than standard de Sitter inflation ($a(t)=\exp(H
t)$), but faster than power law inflation ($a(t)= t^p,\ p>1$). The
intermediate inflationary model was introduced as an exact
solution corresponding to a particular scalar field potential of the type
$V(\phi)\propto \phi^{-4(f^{-1}-1)}$ (in the slow-roll approximation), where $0<f<1$. With this sort of potential it is
possible in the slow-roll approximation to have a spectrum of
density perturbations which presents a scale-invariant spectral
index $n_s=1$, i.e. the so-called Harrison-Zel'dovich spectrum of
density perturbations, provided $f$ takes the value of
2/3 \cite{Barrow2}. Even though this kind of spectrum is disfavored
by the current Wilkinson Microwave Anisotropy Probe (WMAP)
data \cite{B,K,H,D}, the inclusion of tensor perturbations, which
could be present at some point by inflation and parametrized by
the tensor-to-scalar ratio $r$, allows
 the conclusion that $n_s \geq 1$ providing  that the value of $r$ is significantly
nonzero \cite{ratio r}. In fact, in Ref.\cite{Barrow3} was shown
that the combination $n_s=1$ and $r>0$ is given by a version of
the intermediate inflationary scenario in which the scale factor varies as
$a(t)\propto e^{t^{2/3}}$ within the slow-roll approximation.

On the other hand, motivations also coming from string theory,
there has been carried out a less standard theory of gravity,
namely the so called scalar-tensor theory of
gravity \cite{JBD,BEPA,varios3,varios4}. The archetypical theory
associated with scalar-tensor models is the Jordan-Brans-Dicke
(JBD) gravity \cite{JBD}. The JBD theory is a class of model in
which the effective gravitational coupling evolves with time. The
strength of this coupling is determined by a scalar field, the
so-called JBD field, which tends to the value $G^{-1}$. The origin
of JBD theory is in Mach's principle according to which the
property of inertia of material bodies arises from their
interactions with the matter distributed in the universe. In
modern context, JBD theory appears naturally in supergravity
models, Kaluza-Klein theories and in all known effective string
actions \cite{F,AChF,FT1,FT2,CMPF,CKP,GSW}.

In this paper we would like to study intermediate inflationary
universe model in a JBD theory. We will write the Friedmann  field
equations, together with the corresponding scalar field equations
(inflaton and JBD fields). The intermediate inflationary period of
inflation will be consistently described in the slow-roll approximation. Scalar
and tensor perturbations will be expressed in terms of the
parameters that appear in our model, these parameters will
be constrained by taking into account the WMAP five and seven year
data.

The outline of the paper is as follows. The next section presents
the field equations in the Einstein frame. In Section \ref{sect4}
we study the slow-roll approximation. Section \ref{sect5} deals
with the calculations of cosmological scalar perturbations. Then
we describe the quantum generation of fluctuations together with
the spectrum of comoving curvature perturbations in section \ref{sect6}.
Section \ref{sect7} deals with tensor perturbations. Finally, in
Section \ref{conclu} we conclude our findings.

\section{Background Equations in the Einstein Frame}\label{sect3}
A wide class of non-Einstein gravity models can be recast in the action \cite{Starobinsky2}:
\begin{eqnarray}
\label{BDI1}
S=\int \sqrt{-g}d^4x\left[\frac{1}{2\kappa^2}R+\frac{1}{2}g^{\mu\nu}\partial_{\mu}\chi\partial_{\nu}\chi-U(\chi)+\frac{1}{2}e^{-\gamma \kappa \chi}g^{\mu\nu}\partial_{\mu}\phi\partial_{\nu}\phi-e^{-\beta\kappa\chi}V(\phi)\right],
\end{eqnarray}
where $R$ is the Ricci scalar, $\kappa^2=8\pi G$, with $c=\hbar=1$,
$\beta$ and $\gamma$ are constants, $\chi$ and $\phi$ are the
dilaton and inflaton fields, respectively.

The Jordan-Brans-Dicke  action in the Jordan frame is given as:
\begin{eqnarray}
\label{BDI2}
S=\int\sqrt{-\hat{g}}d^4x\left[\frac{\Phi_{BD}}{16\pi}{\cal \hat{R}}+\frac{\omega^2}{16\pi\Phi_{BD}}\hat{g}^{\mu\nu}\partial_{\mu}\Phi_{BD}\partial_{\nu}\Phi_{BD}+\frac{1}{2}\hat{g}^{\mu\nu}\partial_{\mu}\phi\partial_{\nu}\phi-V(\phi)\right],
\end{eqnarray}
and it is recovered by a conformal transformation \cite{Conformal} on the
action (\ref{BDI1}) with the condition $U(\chi)=0$
\begin{eqnarray}
\label{BDI3}
g_{\mu\nu}=\Omega^2\hat{g}_{\mu\nu}\ \ \ \ \ \textrm{and} \ \ \ \ \ \Omega^2\equiv\frac{\kappa^2}{8\pi}\Phi_{BD}\equiv e^{\gamma\kappa\chi},
\end{eqnarray}
for $\beta=2\gamma=\frac{2}{\sqrt{\omega+\frac{3}{2}}}$. $\Phi_{BD}$ is the Brans-Dicke field and $\omega$ is the Brans-Dicke parameter.

Observational measurements \cite{ObsWill,ObsBertotti}
constraint the BD parameter to be very large $\omega\gg1$. On the other hand, the BD field remains very close to a constant after inflation, in the radiation and matter domination eras \cite{JBD}. In order to recover the right value of the newtonian gravitational constant after inflation we will consider $\chi$ equals to zero at the end of the inflationary stage. 

From the action (\ref{BDI1}) in a flat Friedmann-Robertson-Walker
(FRW)  metric, tacking $U(\chi)=0$, we get the following set of field equations:
\begin{eqnarray}
\label{BDI4}
\ddot{\chi}+3H\dot{\chi}+\frac{\gamma\kappa}{2}e^{-\gamma\kappa\chi}\dot{\phi}^2-\beta\kappa e^{-\beta\kappa\chi} V(\phi)&=&0,\\
\label{BDI5}
\ddot{\phi}+3H\dot{\phi}-\gamma\kappa\dot{\chi}\dot{\phi}+e^{(\gamma-\beta)\kappa\chi}V'(\phi)&=&0,\\
\label{BDI6}
3H^2-\kappa^2\left(\frac{1}{2}\dot{\chi}^2+\frac{1}{2}e^{-\gamma\kappa \chi}\dot{\phi}^2+e^{-\beta\kappa\chi}V(\phi)\right)&=&0,
\end{eqnarray}
where overdots denote derivatives with respect to $t$ and a prime
denotes derivative respect to the scalar field $\phi$,
$H\equiv\frac{\dot{a}(t)}{a(t)}$ is the cosmic expansion rate and
$a(t)$ is the scale factor. In the next section we will solve the set of equations Eqs.(\ref{BDI4})-(\ref{BDI6}) in the
slow-roll approximation.

\section{The slow-roll approximation}\label{sect4}
The slow-roll regime of an inflationary era is presented  when
$\left|\frac{\dot{H}}{H^2}\right|\ll1$ and
$\left|\frac{\ddot{H}}{\dot{H}H}\right|\ll1$ \cite{Liddle-Lyth-0}.
These conditions into Eqs.(\ref{BDI4})-(\ref{BDI6}) impose the
constraints:
\begin{eqnarray}
\label{BDI10}
\frac{1}{2}e^{-\gamma\kappa\chi}\dot{\phi}^2+\frac{1}{2}\dot{\chi}^2&\ll& e^{-\beta\kappa\chi}V(\phi),\\
\label{BDI11}
\ddot{\phi}-\gamma\kappa\dot{\chi}\dot{\phi}&\ll& H\dot{\phi},\\
\label{BDI12}
\ddot{\chi}+\frac{1}{2}\gamma\kappa e^{-\gamma\kappa\chi}\dot{\phi}^2&\ll& H\dot{\chi},
\end{eqnarray}
which transform the set of field equations, Eqs.(\ref{BDI4})-(\ref{BDI6}), into the equations:
\begin{eqnarray}
\label{BDI13}
3H\dot{\chi}-\beta\kappa e^{-\beta\kappa\chi} V(\phi)&=&0,\\
\label{BDI14}
3H\dot{\phi}+e^{(\gamma-\beta)\kappa\chi}V'(\phi)&=&0,\\
\label{BDI15}
3H^2-\kappa^2e^{-\beta\kappa\chi}V(\phi)&=&0.
\end{eqnarray}

From this set of differential equations we easily realize that $\chi$ is given by:
\begin{eqnarray}
\label{BDI16}
\chi=\frac{\beta}{\kappa}\ln\left(\frac{a}{a_b}\right)+\chi_b,
\end{eqnarray}
where the subscript $b$ will denote values at the beginning of the inflationary epoch.

By choosing an ansatz for the potential as $V(\phi)=V_0\phi^n$, from Eqs.(\ref{BDI13})-(\ref{BDI15}) we obtain:
\begin{eqnarray}
\label{BDIMi1}
\phi^2-\phi_b^2=\frac{2ne^{\gamma\kappa\chi_b}}{\beta\gamma\kappa^2}\left(1-\left(\frac{a}{a_b}\right)^{\beta\gamma}\right)
\ \ \ \ \ \textrm{and}\ \ \ \ \ 
e^{\gamma\kappa\chi}=e^{\gamma\kappa\chi_b}+\frac{\beta^2\kappa^2f}{16(1-f)}\left(\phi^2-\phi_b^2\right).
\end{eqnarray}

For the specific case of a Jordan-Brans-Dicke theory, where $\gamma=\frac{\beta}{2}$, we find:
\begin{eqnarray}
\label{BDIMi2}
a(t)=a_b\left(\frac{1+\frac{A}{p} t^f}{1+\frac{A}{p} t_b^f}\right)^{p}
\ \ \ \ \ \textrm{and}\ \ \ \ \ 
\phi(t)=\left(\frac{8\sqrt{V_0}(1-f) t}{\sqrt{3}\kappa f^2}\right)^{\frac{f}{2}}
\end{eqnarray}
with $p\equiv\frac{2}{\beta^2}$, $f\equiv\frac{4}{4-n}$,
$A\equiv-\frac{2}{(1+C_2) \beta ^2} \left(-\frac{C_1
C_2}{f}\right)^{f}$, $t_b\equiv-\frac{f}{C_1C_2}$, $C_1\equiv
e^{-\gamma  \kappa  \chi_b}\frac{ \sqrt{V_0} \beta ^2 \kappa
\phi_b^{n/2}}{2 \sqrt{3}}$ and $C_2\equiv-e^{\gamma  \kappa
\chi_b} \frac{16 (1-f)}{f \beta ^2 \kappa ^2\phi_b^2}$.

The parameter $A$ in Eq.(\ref{BDIMi2}) has to be
positive in order to get an increasing scale factor function.
Consequently, we have the following constraints: $C_2<-1$, $0<f<1$
and $n<0$. In this case the potential $V(\phi)$ does not have a
minimum and therefore a nonstandard way of reheating in the
universe is required \cite{Curvaton}. 

From here on it is assumed
$\gamma=\frac{\beta}{2}$ although sometimes $\gamma$ is preserved
to shorten the length of the equations.

We note that the scale factor in Eq.(\ref{BDIMi2}) is a
generalization of the scale factor corresponding to intermediate
inflation in the Einstein theory \cite{Barrow1}, in the case $p\rightarrow\infty$
we recover $a(t)\propto e^{At^f}$. On the other hand, the authors
of Ref.\cite{LaSteinhardt} found the same form for $a(t)$ when
they first studied a cosmological model in a JBD theory, extended inflation.

The number of e-folds between any time $t$ and the beginning of
inflation $t_b$ is:
\begin{eqnarray}
\label{BDIMi4} N \equiv 
\ln\left(\frac{a}{a_b}\right)=p\ln\left(\frac{1+
\frac{A}{p}t^f}{1+\frac{A}{p}t_b^f}\right).
\end{eqnarray}
$N$ is always lower than the number of e-folds that we would get in an
intermediate inflationary model in the Einstein theory
\cite{Barrow-Liddle}. Expression (\ref{BDIMi4}) converges to the
Einstein case in the limit $p\rightarrow\infty$.

It is well known  that an intermediate stage of inflation needs an
additional mechanism to bring inflation to an end
\cite{Barrow3}. We will consider that this mechanism
starts after $N_T$ e-folds since the beginning of inflation. We
normalize $\chi$ in such a way that after $N_T$ e-folds the value
of the  field $\chi$ becomes zero, therefore
$\chi_b=-\frac{\beta}{\kappa}N_T$. We assume that the value of
$\chi$ remains zero after that time in order to fulfill the
condition $\Phi_{BD}=\frac{1}{G}$ after inflation.

It is convenient to calculate the so-called slow-roll parameters:
\begin{eqnarray}
\label{BDIMM2}
\epsilon&\equiv&-\frac{\dot{H}}{H^2}=\frac{\beta^2}{2}+e^{\gamma\kappa\chi}\frac{1}{2\kappa^2}\left(\frac{V'}{V}\right)^2,\\
\label{BDIMM3}
\eta&\equiv&-\frac{\ddot{H}}{\dot{H}H}=-2\epsilon+\frac{7\beta^2}{2}-\frac{3\beta^4}{4\epsilon}+e^{\gamma\kappa\chi}\left(\frac{2}{\kappa^2}-\frac{\beta^2}{\epsilon\kappa^2}\right)\frac{V''}{V},
\end{eqnarray}
which will be useful in the study of the perturbations of the model.
We recall that $\epsilon<1$ implies $\ddot{a}>0$, i.e. it guarantees
the existence of an inflationary period.

From Eqs.(\ref{BDIMi1}) and (\ref{BDIMM2}) we note that $\epsilon$ is a
decreasing function of time. Following Refs.\cite{Barrow3,delCampo}
we assume that the intermediate inflationary era begins at the earliest possible
stage when $\epsilon=1$, which corresponds to:
\begin{eqnarray}
\label{BDIMM4}
t_b=\left(\frac{2(1-f)}{A(2f-\beta^2)}\right)^{\frac{1}{f}},\ \ \ \ \phi_b=\frac{(1-f)}{f\kappa}\frac{4e^{\frac{\beta \kappa \chi_b}{4}}}{\sqrt{2-\beta^2}},\ \ \ \ V_0=\frac{3 A^{\frac{2}{f}} f^4 \kappa ^2 e^{\frac{\beta  \kappa \chi_b}{f}} }{4^{3-\frac{3}{f}}(1-f)^2}\left(\frac{(1-f) \left(2 f-\beta ^2\right)}{f^2 \left(2-\beta ^2\right) \kappa ^2}\right)^{\frac{2}{f}}.
\end{eqnarray}

We note that in the limit $\beta\rightarrow0$ the expressions for
$\phi(t)$ in Eq.(\ref{BDIMi2}), $\epsilon$ in Eq.(\ref{BDIMM2})
and $\eta$ in Eq.(\ref{BDIMM3}) go to the standard slow-roll
relations corresponding to the intermediate inflationary universe
model \cite{Barrow3}, as well as the expression for
$V(\phi)$, where $V_0\rightarrow\frac{3A^2f^2}{\kappa^2}
\left(\frac{8A(1-f)}{f\kappa^2}\right)^{2\left(\frac{1-f}{f}\right)}$.

\begin{figure}[ht!]
\includegraphics[scale=0.5]{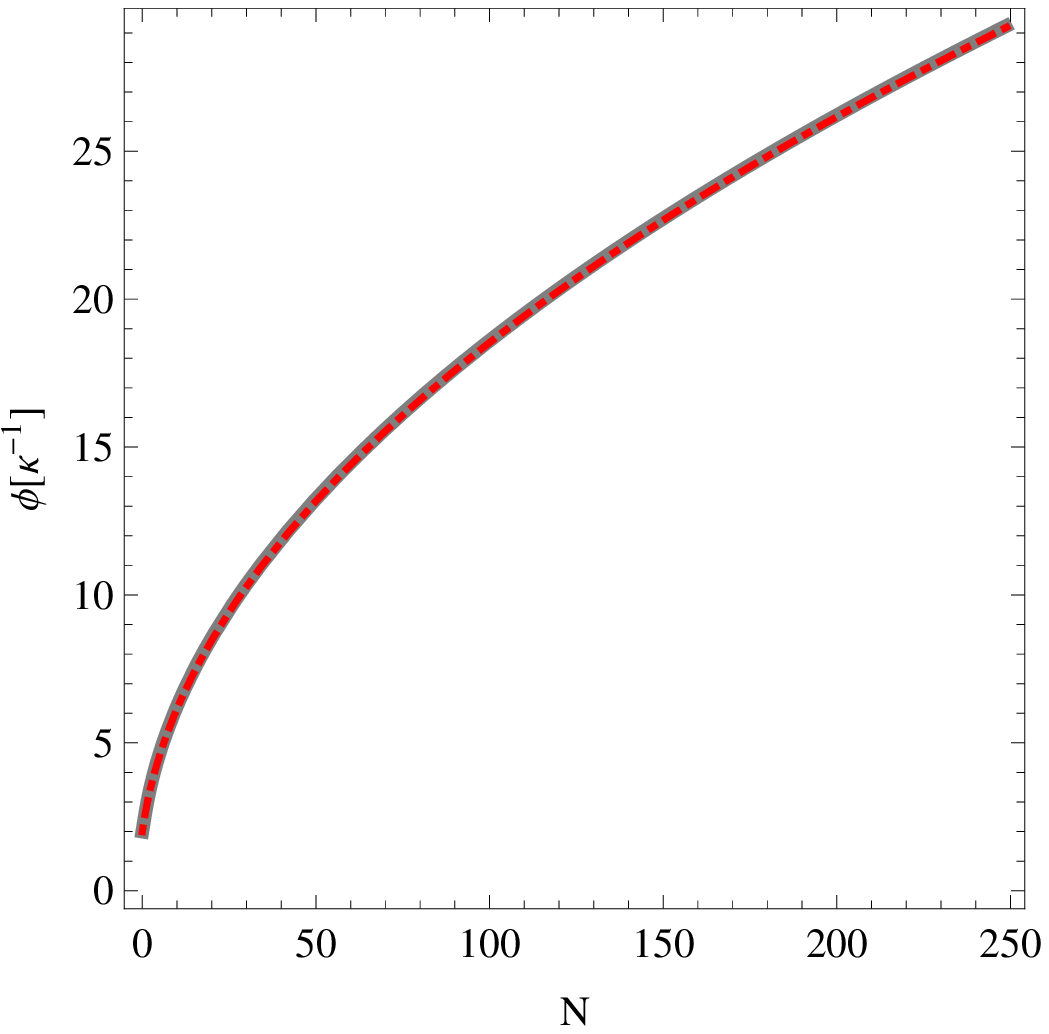}
\hspace{2cm}
\includegraphics[scale=0.52]{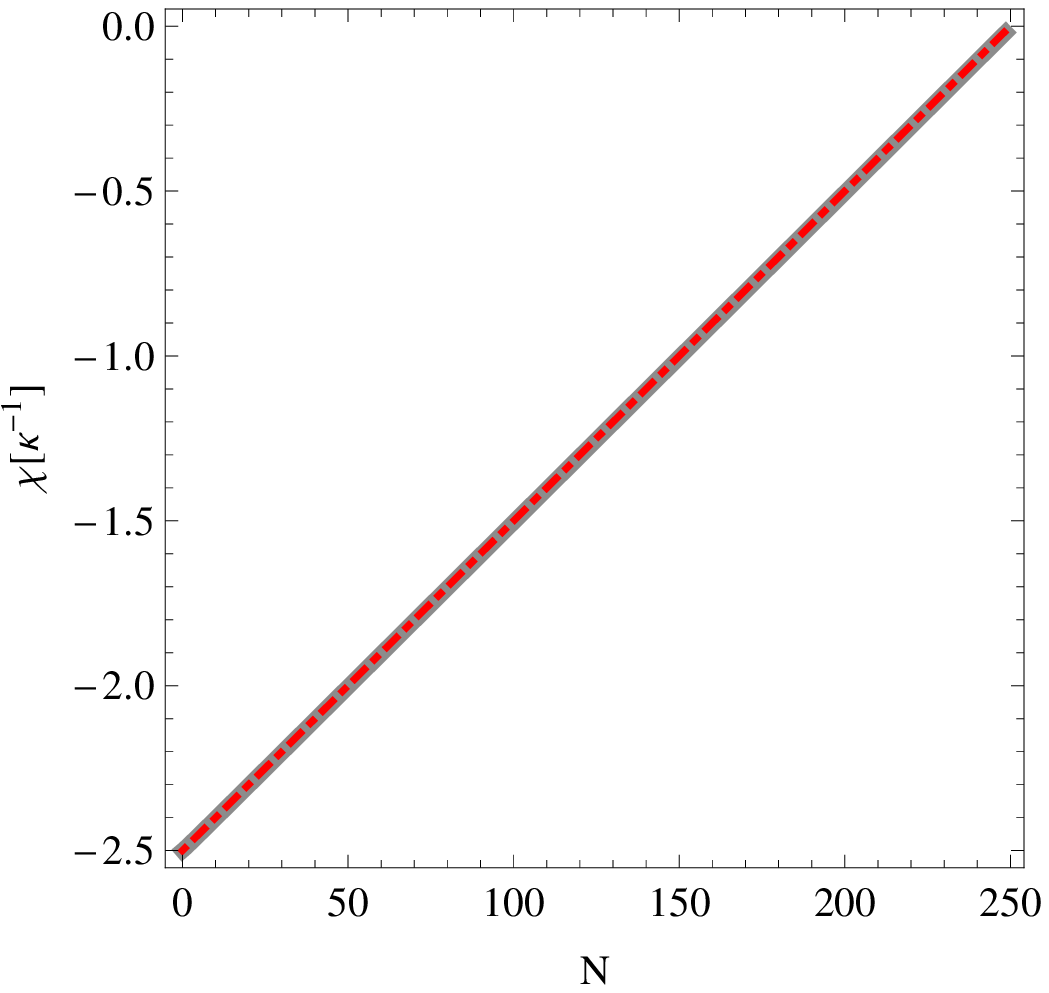}
\caption{\label{figN1} The panels show 250 e-folds of inflation for $f=0.7$ and $\beta=0.01$, the exact solutions to Eqs.(\ref{BDI4})-(\ref{BDI6}) coincides with the solutions for these equations in the slow-roll approximation (dotdashed line). For the last 100 e-folds the error in use the slow-roll approximation is less than 1$\%$.}
\end{figure}

We can check the consistency of the slow-roll approximation numerically. We get the solutions to the Eqs.(\ref{BDI4})-(\ref{BDI6}) and compare with the solutions to Eqs.(\ref{BDI13})-(\ref{BDI15}) (see FIG.\ref{figN1}). The initial values for $\dot{\phi}(t)$ and $\dot{\chi}(t)$ in the exact solution were set from the slow-roll differential equations (\ref{BDI13}) and (\ref{BDI14}). We can see from FIG.\ref{figN1} that the slow-roll approximation for the given parameters is a very good approximation, this is also valid for the other parameters in the considered ranges.

\section{Linear Order Scalar Perturbations}\label{sect5}

We analyze the cosmological scalar perturbations in the longitudinal gauge \cite{Mukhanov}, we
consider the perturbed metric to be:
\begin{eqnarray}
\label{BDI20}
ds^2=(1+2\Phi)dt^2-a(t)^2(1-2\Psi)\delta_{ij}dx^idx^j.
\end{eqnarray}
When we introduce this perturbed metric into the Einstein field
equations with two scalar fields ($\phi$ and $\chi$) we get the
following set of linear order perturbed field equations:
\begin{eqnarray}
&& \Phi=\Psi\label{BDI21}\\
&& \dot{\Phi}+H\Phi=\frac{\kappa^2}{2}\left(\dot{\chi}\delta\chi+e^{-\gamma\kappa\chi}\dot{\phi}\delta\phi\right)\label{BDI22},\\
&&\ddot{\delta\chi}+3H\dot{\delta\chi}+\left(\frac{k^2}{a^2}-\frac{(\gamma\kappa)^2}{2}e^{-\gamma\kappa\chi}\dot{\phi}^2+(\beta\kappa)^2e^{-\beta\kappa\chi}V(\phi)\right)\delta\chi+\gamma\kappa e^{-\gamma\kappa\chi}\dot{\phi}\dot{\delta\phi}\nonumber \\
&&-\beta\kappa e^{-\beta\kappa\chi}V'(\phi)\delta\phi =2(\ddot{\chi}+3H\dot{\chi})\Phi+\dot{\Phi}\dot{\chi}+3\dot{\Psi}\dot{\chi}+\gamma\kappa e^{-\gamma\kappa\chi}\dot{\phi}^2\Phi,\label{BDI23}\\
\textrm{and}\ \ \ \ \ \ \ \ \ \ && \ddot{\delta\phi}+(3H-\gamma\kappa\dot{\chi})\dot{\delta\phi}+\left(\frac{k^2}{a^2}+e^{(\gamma-\beta)\kappa\chi}V''(\phi)\right)\delta\phi-\gamma\kappa\dot{\phi}\dot{\delta\chi}\nonumber\\
&&+(\gamma-\beta)\kappa V'(\phi)e^{(\gamma-\beta)\kappa\chi}\delta\chi=2(\ddot{\phi}+3H\dot{\phi})\Phi+\dot{\Phi}\dot{\phi}+3\dot{\Psi}\dot{\chi}-2\gamma\kappa\dot{\phi}\dot{\chi}\Phi,\label{BDI24}
\end{eqnarray}
where $\delta\phi$ and $\delta\chi$ are gauge invariant
fluctuations of the respective fields and $k$ stands for the
Fourier space decomposition.

We use the slow-roll approximation in
Eqs.(\ref{BDI22})-(\ref{BDI24})  and given that we are interested
in the non-decreasing adiabatic and isocurvature modes on large
scales $k\ll aH$ \cite{Liddle-Lyth}, we can consistently neglect 
the terms containing $\dot{\Phi}$ and those terms containing
second order time derivatives. Under these approximations
Eqs.(\ref{BDI22})-(\ref{BDI24}) reduce to:
\begin{eqnarray}
&&\Phi=\frac{\kappa^2}{2H}\left(\dot{\chi}\delta\chi+e^{-\gamma\kappa\chi}\dot{\phi}\delta\phi\right)=\frac{\beta\kappa}{2}\delta\chi-\frac{V'(\phi)}{2V(\phi)}\delta\phi,\label{BDI25}\\
&&3H\dot{\delta\chi}+(\beta\kappa)^2e^{-\beta\kappa\chi}V(\phi)\delta\chi-\beta\kappa e^{-\beta\kappa\chi}V'(\phi)\delta\phi=2\beta\kappa e^{-\beta\kappa\chi}V(\phi)\Phi,\label{BDI26}\\
\textrm{and}\ \ \ \ \ \ \ \ \ \ &&3H\dot{\delta\phi}+e^{(\gamma-\beta)\kappa\chi}V''(\phi)\delta\phi+(\gamma-\beta)\kappa V'(\phi)e^{(\gamma-\beta)\kappa\chi}\delta\chi=-2e^{(\gamma-\beta)\kappa\chi}V'(\phi)\Phi.\label{BDI27}
\end{eqnarray}

The authors in Ref.\cite{Starobinsky,Chiba-Sugiyama-Yokoyama}
have found a solution to the set of Eqs.(\ref{BDI25})-(\ref{BDI27}):
\begin{eqnarray}
&&\frac{\delta\chi}{\dot{\chi}}=\frac{C_1}{H}-\frac{C_3}{H},\label{BDI28}\\
&&\frac{\delta\phi}{\dot{\phi}}=\frac{C_1}{H}+\frac{C_3}{H}(e^{-\gamma\kappa\chi}-1),\label{BDI29}\\
\textrm{and}\ \ \ \ \ \ \ \ \ \ &&\Phi=-C_1\frac{\dot{H}}{H^2}+C_3\left(\frac{(1-e^{\gamma\kappa\chi})}{2\kappa^2}\left(\frac{V'(\phi)}{V(\phi)}\right)^2-\frac{\beta^2}{2}\right),\label{BDI30}
\end{eqnarray}
where $C_1$ and $C_3$ are two integration constants related with the initial values of $\delta\phi$, $\delta\chi$, $\phi$ and $\chi$. The terms proportional to $C_1$ and $C_3$ represent adiabatic and
isocurvature modes, respectively\cite{Starobinsky2}. The
isocurvature nature of the term proportional to $C_3$ is
guaranteed by the fact that the second term in Eq.(\ref{BDI30}) is
vanishingly small after inflation when $\chi(t)\approx0$.

In order to get the spectrum of scalar perturbations we  introduce
the gauge-invariant quantity named the comoving curvature
perturbation, ${\cal R}$ \cite{Garcia-Bellido-Wands}:
\begin{eqnarray}
\label{BDI31}
{\cal R}=\Psi-\frac{H}{\dot{H}}\left(\dot{\Psi}+H\Phi\right)=-\frac{H^2}{\dot{H}}\Phi,
\end{eqnarray}
where the latter expression is valid  for large scales in the
slow-roll regime. By substituting $\Phi$ given by Eq.(\ref{BDI30})
into Eq.(\ref{BDI31}) we obtain:
\begin{eqnarray}
\label{BDI32}
{\cal R}=C_1-C_3W,
\end{eqnarray}
with $W=1-\left(e^{\gamma\kappa\chi(t)}+\beta^2\kappa^2\left(\frac{V(\phi)}{V'(\phi)}\right)^2\right)^{-1}$.

As we see from Eq.(\ref{BDI32}), the term $W$ is responsible for
the change of ${\cal R}$ during the inflationary stage. For
intermediate inflation in a JBD theory we can calculate $W$ by
using Eqs.(\ref{BDI16}), (\ref{BDIMi1}) and (\ref{BDIMi4}) as well
as the definition of $V(\phi)$:
\begin{eqnarray}
\label{BDI33}
W(N)=1+\frac{e^{\frac{N_T\beta^2}{2}}(1-f)(2-\beta^2)}{2f-\beta^2-e^{\frac{N\beta^2}{2}}(2-\beta^2)},
\end{eqnarray}
where $N_T$ is the total amount of e-folds of inflation for which
the value of $\chi$ goes to zero. Here we have used the number of
e-folds $N$ to describe time evolution because it is more
convenient in the subsequent analysis.

We note that the case $\beta=0$ corresponds to $W=0$, which is
expected because when the BD parameter is zero there is only one scalar field driving
inflation, and in that case the comoving curvature perturbation
$\cal{R}$ remains constant on large scales \cite{Mukhanov}.

\begin{figure}[ht!]
\centering
\includegraphics[scale=0.8]{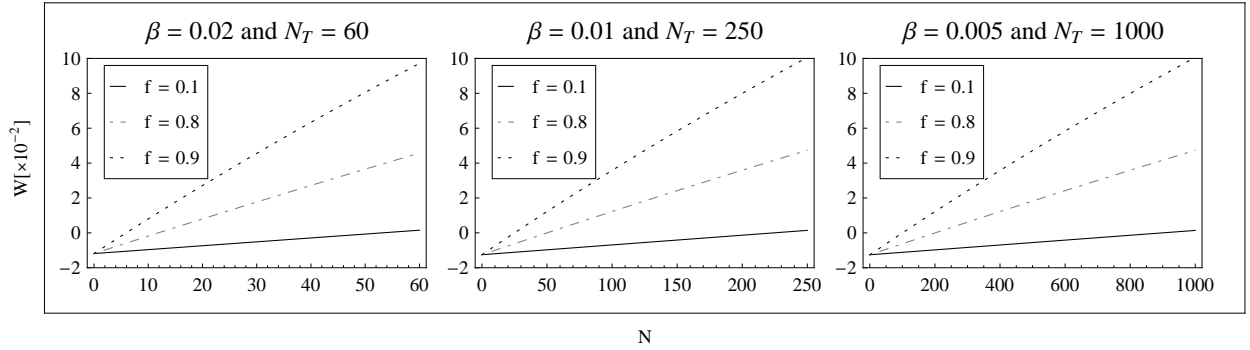}
\caption{The variation of the function $W$ during the intermediate inflationary stage for different values of $\beta$ and $f$. In order to have $\vert{W}\vert<0.1$ the maximum number of e-folds allowed $N_T$ changes depending on $\beta$ for $0<f<1$.}
\label{fig1}
\end{figure}

The current observational constraints bring $\beta$ to an upper
limit given by $\beta \le0.02$ \cite{ObsBertotti}. We see from
FIG. \ref{fig1} that for a wide range of $N_T$ we can find values
for $\beta$ and $f$ in the allowed ranges in such a way that
$\vert W\vert<0.1$. For example, for $\beta=0.01$, $f=0.8$ and
$N_T\le250$ we get $\vert W\vert<0.05$.

Given that the constants $C_1$ and $C_3$ are related to the initial values of the perturbed fields it is expected they to be of the same order \cite{Starobinsky2}, then to impose $\vert W\vert<0.1$
 guarantees that the variation of ${\cal R}$ during inflation due to the presence
 of isocurvature perturbations is small, i.e. for a given $\beta$ and $f$ we
 can find a maximum $N_T$ value for which $\vert W\vert<$ desired value.

In the following we will consider for the comoving curvature
perturbation  ${\cal R}\approx C_1$, i.e. ${\cal R}$ remains
constant after a given scale $k$ leaves the Hubble horizon during
inflation. Furthermore, we will assume that the mechanism which is
needed to properly finish inflation is not going to modify the
results for ${\cal R}$.

\begin{figure}[ht!]
\includegraphics[scale=1]{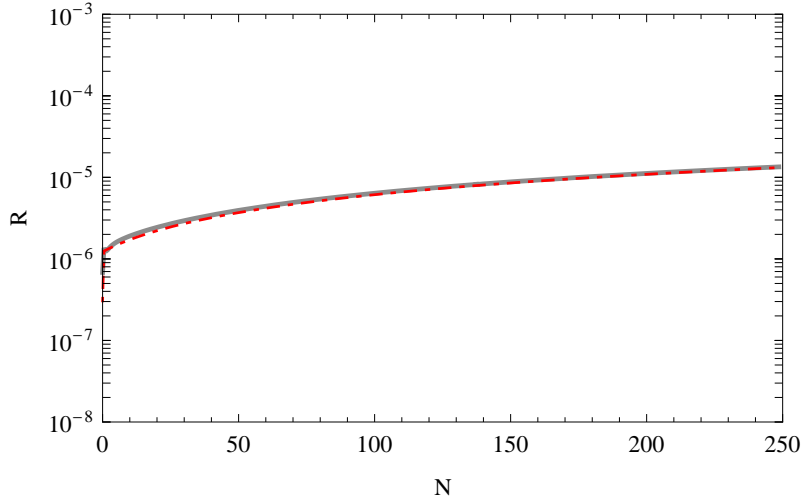}
\caption{\label{FigN2} The figure shows a comparison of the comoving curvature perturbation between the solutions to the exact perturbed field equations and the solutions to these equations in the slow-roll approximation. For the last 100 e-folds the error in use the slow-roll approximation is less than 5$\%$. The values for the parameters and the initial conditions are the same considered in FIG.\ref{figN1}. We have normalized to get $P_{\cal R}=2.3\times10^{-9}$ at the crossing time when $N_*=50$ e-folds.}
\end{figure}

In FIG.\ref{FigN2} we plot the comoving curvature perturbation, the exact solution and the solution in the slow-roll approximation are compared.
We confirmed, for the allowed range of parameters for $\beta$, $f$ and $N_T$, that the slow-roll approximation is adequated when we set the initial values of $\dot{\phi}$, $\dot{\chi}$, $\dot{\delta{\phi}}$ and $\dot{\delta\chi}$ from the equations in the slow-roll approximation.

\section{Quantum Generation of Fluctuations and Spectrum of Curvature
Perturbation}\label{sect6}

The value of the constant $C_1$ in the slow-roll approximation and
for large scales ($k\ll aH$) is gotten from Eqs.(\ref{BDI28}) and
(\ref{BDI29}):
\begin{eqnarray}
\label{BDI34}
C_1=He^{\gamma\kappa\chi}\left(\frac{\delta\chi}{\dot{\chi}}\left(e^{-\gamma\kappa\chi}-1\right)+\frac{\delta\phi}{\dot{\phi}}\right).
\end{eqnarray}

The expectation values of the scalar field perturbations
$\delta\phi$  and $\delta\chi$ are given by random gaussian
variables when they cross outside the Hubble radius ($k\approx
a_*H_*$) \cite{Mukhanov}, these are given by:
\begin{eqnarray}
\label{BDI35}
\langle\vert\delta\phi_*\vert^2\rangle=\frac{H_*^2}{2k^3}e^{\gamma\kappa\chi_*}\ \ \ \ \ \textrm{and}\ \ \ \ \ \langle\vert\delta\chi_*\vert^2\rangle=\frac{H_*^2}{2k^3}
\end{eqnarray}
respectively, here the subscript $*$ denotes the crossing time and  the brackets
the expectation value of the respective random variable.

The spectrum of the comoving curvature perturbation ${\cal R}$  is
defined by \cite{Garcia-Bellido-Wands}:
\begin{eqnarray}
\label{BDI36}
P_{{\cal R}}(k)=\frac{4\pi k^3}{(2\pi)^3}\langle \vert {\cal R} \vert^2\rangle.
\end{eqnarray}

From the discussion in Section \ref{sect5} and Eqs.(\ref{BDI34})
and (\ref{BDI35}) we obtain:
\begin{eqnarray}
\label{BDI37}
P_{{\cal R}}(k_*)=\frac{4\pi k_*^3}{(2\pi)^3}\langle \vert C_1 \vert^2\rangle=\left[\frac{H^2e^{2\gamma\kappa\chi}}{(2\pi)^2}\left(\left(e^{-\gamma\kappa\chi}-1\right)^2\frac{H^2}{\dot{\chi}^2}+\frac{H^2e^{\gamma\kappa\chi}}{\dot{\phi}^2}\right)\right]_{t_*}.
\end{eqnarray}
We note that the presence of a second scalar field during inflation modifies the form of the standard spectrum \cite{Starobinsky2}, the standard form \cite{Linde} is recovered when we take
the limit $\gamma\rightarrow0$. In terms of $\chi$ and the
parameters of the model, the spectrum becomes:
\begin{eqnarray}
\label{BDIMMM1}
P_{{\cal R}}(k_*)=\frac{\kappa^2}{12}V_0\phi_*^{n}\left(\left(1-e^{-\frac{1}{2}\beta\kappa\chi_*}\right)^2\frac{\kappa^2}{\beta^2}+e^{-\frac{1}{2}\beta\kappa\chi_*}\frac{\phi_*^2}{n^2}\right),
\end{eqnarray}
where $n=4\left(\frac{f-1}{f}\right)$, $\phi_*$ is given by
Eq.(\ref{BDIMi1}) in terms of $\chi_*$ and $\chi_*$ is related to $N_*$ by Eq.(\ref{BDI16}) where $\chi_b=-\frac{\beta}{\kappa}N_T$.

\begin{figure}[ht!]
\centering
\includegraphics[scale=0.8]{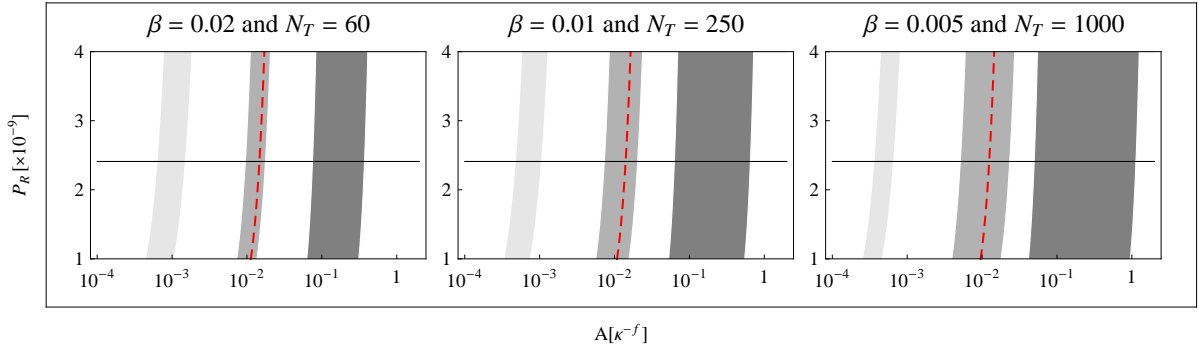}
\caption{The panels show the value of $A$ in units $\kappa^{-f}$. The stripes in each panel represents from darkest to lightest $f=0.4,0.6,0.8$. The dashed line corresponds to a scale $k=0.002\textrm{Mpc}^{-1}$ leaving the horizon after 10 e-folds since the beginning of inflation for different sets of parameters.}
\label{fig3}
\end{figure}

From the seven-year WMAP data we know that the amplitude  of the
spectrum is $P_{\cal R}(k_*)=2.43\times10^{-9}$ for a scale
$k_*=0.002\textrm{Mpc}^{-1}$ \cite{WMAP}. By imposing this constraint in Eq.(\ref{BDIMMM1}) we
can get the value of the constant $V_0$ for given values of the
parameters $\beta$, $f$, $N_T$ and $N_*$. In FIG. \ref{fig3} we show the value of the parameter $A$, related to $V_0$ through Eq.(\ref{BDIMM4}).

The scale dependence of the spectrum is characterized by the
spectral index $n_s(k)$ whereas the scale dependence of the
spectral index is given by the running $\alpha_s$
\cite{Liddle-Lyth-0}.

We can calculate the spectral index and the running for scalar
perturbations  as:
\begin{eqnarray}
\label{BDMix}
n_s(k_*)&\equiv&1+\frac{d\ln P_{{\cal R}}}{d\ln k}=1-4\epsilon+\eta+\frac{\beta ^2}{2} Z_1,\\
\alpha_s(k_*)&\equiv&\frac{d n_s}{d\ln k}=-8\epsilon^2+5\epsilon\eta-\xi^2+\frac{\beta ^4}{4} Z_2,
\label{BDIMM7}
\end{eqnarray}
where we have used $d\ln k=dN$ and
\begin{eqnarray*}
Z&\equiv&e^{\frac{N \beta ^2}{2}}\left(2-\beta ^2\right),\ \ \ \ \ \ \ \ \ Z_0\equiv e^{\frac{N_T \beta ^2}{2}}\left(2-\beta ^2\right),\ \ \ \ \ \ \ \ \ \xi^2\equiv\epsilon\eta-\frac{\sqrt{2\epsilon}}{\kappa}\frac{d\eta}{d\phi}\sqrt{\frac{-2f+Z+\beta^2}{Z_0(1-f)}},\\
Z_1&\equiv&1-\frac{f Z}{f (Z-2)+\beta ^2}+\frac{Z}{Z-2 f+\beta ^2}+\frac{Z^2+(f-1)Z_0^2}{(Z-Z_0)^2+f \left(2 Z (Z_0-1)-Z_0^2\right)+Z \beta ^2},\\
Z_2&\equiv&-1+\frac{f^2 Z^2}{\left(f (Z-2)+\beta ^2\right)^2}+\frac{f Z}{(1-f) \left(f (Z-2)+\beta ^2\right)}-\frac{Z}{(1-f) \left(-2 f+Z+\beta ^2\right)}\\
&-&\frac{\left(Z^2+(f-1)Z_0^2\right)^2}{\left((Z-Z_0)^2+f \left(2 Z (Z_0-1)-Z_0^2\right)+Z \beta ^2\right)^2}+\frac{Z^2-(f-1) Z_0^2}{(Z-Z_0)^2+f \left(2 Z (Z_0-1)-Z_0^2\right)+Z \beta ^2}.
\end{eqnarray*}

We note that both Eqs.(\ref{BDMix}) and (\ref{BDIMM7}) reduce to the result in the Einstein theory when
$\beta\rightarrow0$ \cite{Barrow3}.

\section{Tensor Perturbations}\label{sect7}

In addition to the scalar curvature perturbation, tensor
perturbations can also be generated from quantum fluctuations
during inflation \cite{Mukhanov}. The tensor perturbations do not
couple to matter and consequently they are only determined by the
dynamics of the background metric, so the standard results for the
evolution of tensor perturbations of the metric remains valid. The
two independent polarizations evolve like minimally coupled
massless fields with spectrum \cite{Mukhanov}:
\begin{eqnarray}
\label{BDIMM8}
P_{{\cal T}}(k_*)=8\kappa^2\left(\frac{H}{2\pi}\right)^2_{t_*}.
\end{eqnarray}

From Eq.(\ref{BDI37}) and Eq.(\ref{BDIMM8}) we can determine the
tensor to scalar ratio $r$:
\begin{eqnarray}
r(k_*)=\frac{P_{{\cal T}}}{P_{{\cal R}}}=8\kappa^2\left
[e^{2\gamma\kappa\chi}\left(\left(e^{-\gamma\kappa\chi}-1
\right)^2\frac{H^2}{\dot{\chi}^2}+\frac{H^2e^{\gamma\kappa\chi}}
{\dot{\phi}^2}\right)\right]_{t_*}^{-1},
\end{eqnarray}
which in terms of $Z$ is rewritten as:
\begin{eqnarray}
\label{EqM1}
r(k_*)=\frac{8 (1-f) Z_0^2 \beta ^2}{(Z_*-Z_0)^2+f \left(2 Z_* (-1+Z_0)-Z_0^2\right)+Z_* \beta ^2},
\end{eqnarray}
and it is reduced to $16\epsilon$ in the limit $\beta\rightarrow0$
consistent with Ref.\cite{Barrow3}.

Our analysis has been done in the Einstein frame but the  physical
results have to be interpreted in the Jordan physical frame. The
authors of Ref.\cite{Starobinsky2} analyze this issue and they
conclude that both frames are equivalent given that the JBD field
varies extremely slowly in the post-inflationary universe, then
the adiabatic fluctuations and the tensor perturbations are
described by the same formula in both frames.

\begin{figure}[ht!]
\centering
\includegraphics[scale=0.75]{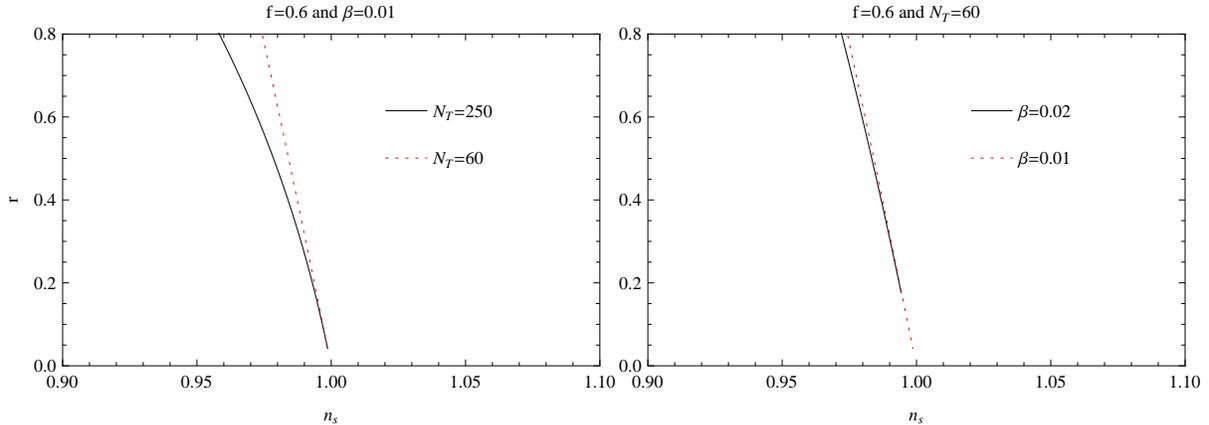}
\caption{Trajectories for different values of the parameter $\beta$ and $N_T$ in the $n_s-r$ plane. Here $N_*$ is a parameter which defines the curve}
\label{fig4}
\end{figure}

As we can see from Eqs.(\ref{BDMix}) and (\ref{EqM1}) $n_s$ and
$r$ only depends on the parameters $f$, $N_T$, $\beta$ and $N_*$. FIG.
\ref{fig4} shows the behavior for the curve $r(n_s)$ for several
choices of the parameters. When we take different values for the
total amount of inflation for a given $\beta$ and $f$, the curves
in the plane $n_s-r$ have different curvature but they have the
same origin in the bottom of the plot. When we take the same
amount of total inflation but different values for $\beta$ for a
given $f$, the curves do not have the same origin.

\begin{figure}[ht!]
\centering
\includegraphics[scale=0.75]{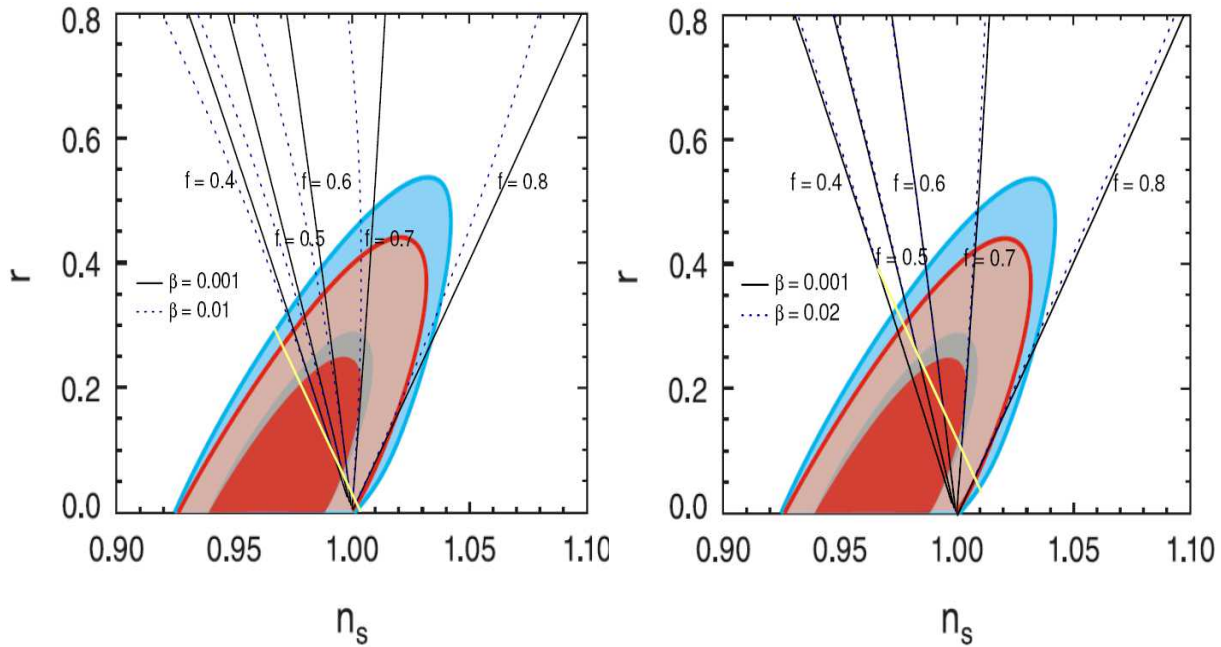}
\caption{Trajectories for different values of the parameter $\beta$ and $f$ in the $n_s-r$ plane. We compare with the WMAP data (five and seven years). The two contours correspond to the $68\%$ and $95\%$ levels of confidence \cite{WMAP2}. The left panel shows $N_T=250$ e-folds for $\beta=0.01$ whereas the right panel is for $N_T=60$ e-folds for $\beta=0.02$.}
\label{fig2}
\end{figure}

FIG. \ref{fig2} shows the dependence of the tensor  to scalar
ratio on the spectral index for different values of the parameters
$\beta$, $f$ and the corresponding maximum $N_T$. We should note from Eqs.(\ref{BDMix}) and (\ref{EqM1}) that these curves are parametrized by the parameter $N_*$. For
$\beta\le0.001$ there is not significant difference with the
predictions in the case of intermediate inflation in the Einstein
theory \cite{Barrow3}. The only difference seems to be
the improvement in the observational constraints to the allowed
plane $n_s-r$ since WMAP3 to WMAP7.

For $\beta=0.01$, $0.4\le f<0.8$ is well supported by  the data.
But there exist a theoretical limit for the model in the maximum
number of e-folds of inflation allowed (represented by the yellow
line in the left panel of FIG. \ref{fig2}), $N_T\le250$.

In the case $\beta=0.02$, the maximum value for $\beta$
constrained from tests of general relativity
\cite{ObsWill,ObsBertotti}, we can have at most 60 e-folds of
inflation  in order to have a comoving curvature perturbation
close to a constant. This constraint in the number of e-folds
exclude $f=0.4$ to be supported by the data  but it still allow
$0.5\le f<0.8$ to be well supported.

We see from FIG. \ref{fig2} that the curve $r=r(n_s)$  enters the
$95\%$ confidence region for $r\le0.41$ which in terms of the
number of e-folds (at the time when a given scale leaves the
horizon) means $N>78$ for $f=0.4$, $N>48$ for $f=0.5$, $N>29$ for
$f=0.6$ and $N>16$ for $f=0.7$. There are not significant
differences for the values of $\beta$ considered in FIG.
\ref{fig2}. On the other hand, we have to consider at least 50
e-folds of inflation to push the perturbations to observable
scales \cite{Liddle-Lyth-2}, which seems to exclude models for
$\beta=0.02$ and $f<0.7$.

\section{Conclusion}\label{conclu}
We have studied in detail the intermediate inflationary scenario
in the context of a JBD theory. This study was realized in the
Einstein frame, but the physical results have to be interpreted in
the Jordan physical frame. In this respect, it has been considered that
both frames are equivalent, providing that the JBD field varies
extremely slowly in the post-inflationary stage of the universe \cite{Starobinsky2}.
In this way, the adiabatic fluctuations and the tensor
perturbations are described equally in both frames. This allows to
obtain explicit expressions for the corresponding power spectrum
of the curvature perturbations $P_{{\cal R}}$, tensor perturbation
$P_{{\cal T}}$, tensor-scalar ratio $r$, scalar spectral index
$n_s$, and its running $\alpha_s$.

In this work the aim has been to study which set of parameters $\beta$, $f$ and $N_T$ allow us to get a dominant contribution of the adiabatic mode to the power spectrum of scalar perturbations. In order to do that we have restricted the maximum
number of e-folds allowed by the model, $N_T$, for a given set of parameters $\beta$ and $f$. For a given value of $\beta$ and $0<f<1$ we get ${\cal R}\approx C_1$ constant for a specific value of $N_T$ provided the desired precision. On the other hand, we have restricted ourselves to $\beta\le0.02$ in light of the previous observational constraints on $\beta$ \cite{ObsWill,ObsBertotti}.

We had checked numerically that the slow-roll approximation is adequated, even to the analysis of first order perturbations, this is valid for the range of parameters considered in this work.

In order to bring some explicit results we have taken the
constraint in the $n_s-r$ plane coming from the seven-year WMAP
data. We have found that the parameter $f$, which initially lies in the
range $0<f<1$ for this model, is well supported by the data as could be
seen from FIG.\ref{fig2}. However, the value of $f$ depends on
the choice of $\beta$ and $N_T$ (FIG.\ref{fig4}). For instance, for the cases
$\beta=0.01$ we have found that $0.4<f<0.8$ is well supported by
the data constrained to the theoretical limit for the model in
the maximum number of e-folds of inflation allowed (represented by
the yellow line in the left panel of FIG. \ref{fig2}),
$N_T\le250$. We also see from FIG.\ref{fig2},
that our values, represented by the curve $r=r(n_s)$, enters the
$95\%$ confidence region for $r\le0.41$, which in terms of the
number of e-folds (at the time when a given scale leaves the
horizon) means $N>78$ for $f=0.4$, $N>48$ for $f=0.5$, $N>29$ for
$f=0.6$ and $N>16$ for $f=0.7$ where there are not significant
differences for the values of $\beta$ considered in the figure.

On the other hand we have to consider at least 50 e-folds of inflation
to push the perturbations to observable scales
\cite{Liddle-Lyth-2}, which seems to exclude models for
$\beta=0.02$ and $f<0.7$. Thus, we see that our study has allowed
us to put restrictions on the parameters that appear in our model by comparing to the WMAP7 results in terms of $n_s-r$ plane.
We have not considered in this work the incidence of the running of the spectral index in the constraints of the model.

Finally in this work, we have not addressed the phenomena of
reheating and possible transition to the standard cosmological
scenario.
A possible calculation for the reheating temperature would give
new constraints on the parameters of our model. We hope to return
to this point in the near future.
\begin{acknowledgments}
This work was supported by the COMISION NACIONAL DE CIENCIAS Y
TECNOLOGIA through FONDECYT Grant N$^{0}$ 1070306 (SdC) and also
was partially supported by PUCV Grant N$^0$ 123.787/2007 (SdC).
M.A.C. was supported by Conicyt and Mecesup.

\end{acknowledgments}

\end{document}